\documentclass[twocolumn,showpacs,preprintnumbers,amsmath,amssymb]{revtex4}
\usepackage{graphicx}
\usepackage{float}
\usepackage{hyperref}
\usepackage{amsmath,amssymb}
\usepackage{bm}
\usepackage{color}

\begin{document}
\title{Topologically equivalent yet radiatively distinct orbits in EMRI system}
\author{Chao-Hui Wang$^{a,b}$}
\author{Shao-Wen Wei$^{a,b}$}
\email{weishw@lzu.edu.cn}
\author{Tao Zhu$^{c}$}
\email{zhut05@zjut.edu.cn}
\author{Yu-Xiao Liu$^{a,b}$}
\email{liuyx@lzu.edu.cn}
\affiliation{
$^{a}$Key Laboratory of Theoretical Physics of Gansu Province, Key Laboratory of Quantum Theory and Applications of MoE, Gansu Provincial Research Center for Basic Disciplines of Quantum Physics,
Lanzhou University, Lanzhou 730000, People's Republic of China.\\
$^{b}$Institute of Theoretical Physics $\&$ Research Center of Gravitation, Lanzhou Center for Theoretical Physics, School of Physical Science and Technology, Lanzhou University, Lanzhou 730000, People's Republic of China.\\
$^{c}$Institute for Theoretical Physics and Cosmology, Zhejiang University of Technology, United Center for Gravitational Wave Physics (UCGWP), Zhejiang University of Technology, Hangzhou, 310023, People's Republic of China.
}

\date{\today}

\begin{abstract}
Multiple potential wells for massive test particles, allowing distinct families of bound orbits to coexist, are a characteristic feature of certain exotic compact objects beyond general relativity. Taking the dyonic black hole as a representative example, we demonstrate that such multi-well geometries generically support multiple coexisting branches of bound orbits, in contrast to the single-branch behavior observed in the Schwarzschild spacetime. Crucially, the periodic orbits sharing identical rational rotation number, and hence identical topological indices can nevertheless produce \emph{radiatively distinct} gravitational waves in a representative extreme-mass-ratio inspirals: their amplitude modulation and harmonic content differ because each branch spans different regions of spacetime curvature. These ``topologically equivalent yet waveform-distinguishable'' signatures provide a direct observational probe of strong field gravitational dynamics beyond general relativity, potentially accessible to future space-based gravitational wave detectors.
\end{abstract}

\maketitle
{\textbf{Introduction}.}---
Black holes predicted by general relativity (GR) provide fundamental laboratories for probing gravity in the strong field region.
The direct detections of gravitational waves from compact binary mergers by LIGO/Virgo~\cite{LIGOScientific:2016aoc,LIGOScientific:2016sjg} and the horizon-scale imaging of M87* by the Event Horizon Telescope~\cite{EventHorizonTelescope:2019dse} have opened unprecedented observational access to this regime.
Forthcoming space-based detectors, such as LISA~\cite{LISA:2017pwj}, Taiji~\cite{Hu:2017mde}, and TianQin~\cite{TianQin:2015yph}, will further probe extreme-mass-ratio inspirals (EMRIs), whose gravitational waves encode detailed information about the underlying geodesic structure of spacetime~\cite{Gair:2004iv,Healy:2009zm,Wang:2025hla,Tu:2023xab,Lu:2025cxx,Haroon:2025rzx}.

In standard GR spacetimes such as Schwarzschild and Kerr, the radial effective potential supports only one single family of bound orbits.
By contrast, a variety of exotic compact objects, including scalarized black holes~\cite{Gan:2021pwu}, hairy black holes~\cite{Gan:2021xdl, Guo:2022muy, Meng:2023htc, Wang:2025hzu}, nonlinear electrodynamics black holes~\cite{Liu:2019rib, Hui:2025ane, Gao:2021kvr, Wei:2020rbh}, wormholes~\cite{Shaikh:2018oul, Xavier:2024iwr}, and other ultra-compact objects~\cite{Cunha:2017qtt}, can develop multi-well effective potentials and multiple photon spheres.
Such structures enrich geodesic dynamics~\cite{Ye:2023gmk,Wei:2022mzv,Delgado:2021jxd}, yet their imprints on gravitational wave signals remain largely unexplored, leaving open whether such multi-well geometries admit observable gravitational wave discriminants.

In this Letter, we take the dyonic black holes governed by quasi-topological nonlinear electrodynamics~\cite{Liu:2019rib} as a representative setting to demonstrate that multi-well effective potentials generically support \emph{multi-branch periodic orbits}, namely distinct families of bound trajectories coexisting at fixed angular momentum.
While such multi-branch structures are not unique to dyonic black holes, we show that their physical implications are far from trivial: even when different branches are classified by the same rational rotation number \(q\) and thus belong to the same topological class that they can emit gravitational waves with sharply distinct amplitude modulations and harmonic content.
These waveform differences persist within the kludge approximation and lie within the sensitivity band of future space-based detectors, rendering topologically equivalent orbits radiatively distinguishable.
This establishes multi-branch
orbits system as \emph{topologically silent yet waveform-distinguishable}, and identifies gravitational waves as a direct observational probe of multi-well strong field geometries and nonlinear electromagnetic interactions.

\begin{figure}[htbp]
\includegraphics[width=0.47\textwidth]{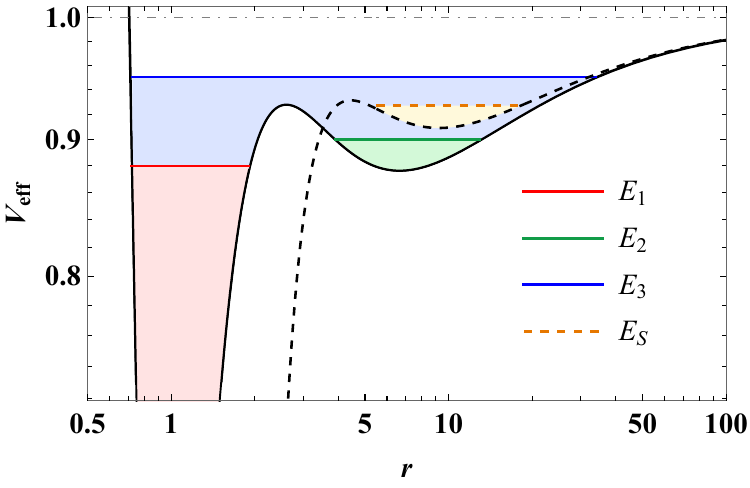}
\caption{Plot of the effective radial potential \(V_{\rm eff}(r)\).
Solid black curve is for the dyonic background showing three nested potential wells (shaded in red, green, and blue colors) with \(L=3.056459521\). Dashed curve is for the Schwarzschild comparison with \(L=3.7\), exhibiting a single well.
}
\label{fig:potential}
\end{figure}

{\textbf{Multi-branch periodic orbits}.}---The dyonic black hole provides a representative example of spacetimes admitting multi-branch periodic orbits while remaining analytically tractable. The corresponding gravitational and electromagnetic dynamics are governed by the Lagrangian~\cite{Liu:2019rib}
\begin{equation}\label{eq:metric}
\mathcal{L} = \sqrt{-g} \Bigl( R - \alpha_1 F^2 - \alpha_2 \bigl[ (F^{(2)})^2 - 2 F^{(4)} \bigr] \Bigr),
\end{equation}
where $\alpha_1$ and $\alpha_2$ are the linear Maxwell and nonlinear electromagnetic coupling coefficients, respectively. The electromagnetic terms are $F^2=F^{\mu\nu}F_{\mu\nu}$, $F^{(2)}=F^{\mu}_{\;\;\nu}F^{\nu}_{\mu}$, and $F^{(4)}=F^{\mu}_{\;\;\nu}F^{\nu}_{\;\;\rho}F^{\rho}_{\;\;\sigma}F^{\sigma}_{\;\;\mu}$. Solving the field equations yields a static, spherically symmetric dyonic black hole with the line element~\cite{Liu:2019rib}
\begin{equation}
\label{eq:metric}
ds^2 = -f(r)\, dt^2 + \frac{dr^2}{f(r)} + r^2 \bigl( d\theta^2 + \sin^2\theta\, d\phi^2 \bigr),
\end{equation}
where the metric function $f(r)$ is given by
\begin{equation}
\label{eq:f}
f(r) = 1 - \frac{2M}{r} + \frac{\alpha_1 p^2}{r^2}
+ \frac{Q^2}{\alpha_1 r^2}\,{}_2F_1\!\left[\tfrac{1}{4},1;\tfrac{5}{4};-\tfrac{4p^2\alpha_2}{r^4\alpha_1}\right].
\end{equation}
Here $M$, $Q$, and $p$ denote the mass, electric charge, and magnetic charge of the black hole, respectively.
The Schwarzschild solution is recovered by setting $Q=p=0$. The spacetime admits two Killing vectors, $\xi^{\mu}=(\partial_t)^{\mu}$ and $\psi^{\mu}=(\partial_\phi)^{\mu}$, associated with the energy $E$ and angular momentum $L$,
\begin{eqnarray}
 -E&=&g_{\mu\nu}u^{\mu}\xi^{\nu}=g_{tt}\dot{t}, \label{conservedE}\\
 L&=&g_{\mu\nu}u^{\mu}\psi^{\nu}=g_{\phi\phi}\dot{\phi}, \label{conservedL}
\end{eqnarray}
where $u^\mu$ is the four-velocity and the dots denote the derivatives with respect to the affine parameter $\tau$.
Imposing the normalization condition $g_{\mu\nu}\dot{x}^{\mu}\dot{x}^{\nu}=-1$, the radial motion reduces to
\begin{eqnarray}
 &\dot{r}^2+V_{\mathrm{eff}}=E^2,\label{rdot}\\
 &V_{\mathrm{eff}}=f(r)\!\left(1+\frac{L^2}{r^2}\right).
\end{eqnarray}
For the Schwarzschild black hole, $V_{\mathrm{eff}}$ admits at most one single potential well (see Fig.~\ref{fig:potential}), allowing only one family of bound orbits at fixed $L$. We work throughout in geometrized units \(G=c=1\) and measure all length scales in units of the black hole mass by setting \(M=1\). In contrast, for the dyonic black hole with parameters
$\alpha_1=1.0494625$, $\alpha_2=2.76$, $p=0.15$, and $Q=1.05$, the effective potential develops a multi-well structure.
Each well can independently bound the motion of particle, giving rise to two distinct bound branches with energies $E_1$ and $E_2$, as well as an additional extended branch with higher energy $E_3$ that spans both wells.

On the equatorial plane, bound motion is characterized by two fundamental frequencies: the radial frequency $\omega_r$ and the azimuthal frequency $\omega_\phi$.
Periodic orbits arise when their ratio is rational.
Following the standard definition~\cite{Levin:2008mq}, we introduce the parameter $q$
\begin{equation}\label{eq:qdef}
q \equiv \frac{\omega_\phi}{\omega_r}-1 = \frac{\Delta\phi_r}{2\pi}-1,
\end{equation}
where $\Delta\phi_r$ is the total azimuthal advance accumulated during one complete radial oscillation.

When $q$ is rational, the orbit closes after $z$ radial oscillations and $w$ whirls, with an additional vertex shift $v$, allowing a discrete topological classification $(z,w,v)$~\cite{Levin:2008mq}.
The total number of azimuthal revolutions in one full periodic cycle is~\cite{Wang:2025wob}
\begin{equation}\label{loop}
n = \frac{\Delta \phi}{2\pi} = z\left(1+w+\frac{v}{z}\right)=z(1+q).
\end{equation}
Consequently, orbits sharing the same rational $q$ and $z$ have identical winding number $n$ and thus the same zoom-whirl-vertex topology, independent of their energy or radial localization.
This provides a precise criterion for topological equivalence among distinct orbital branches.

\begin{figure}[htbp]
\includegraphics[width=0.47\textwidth]{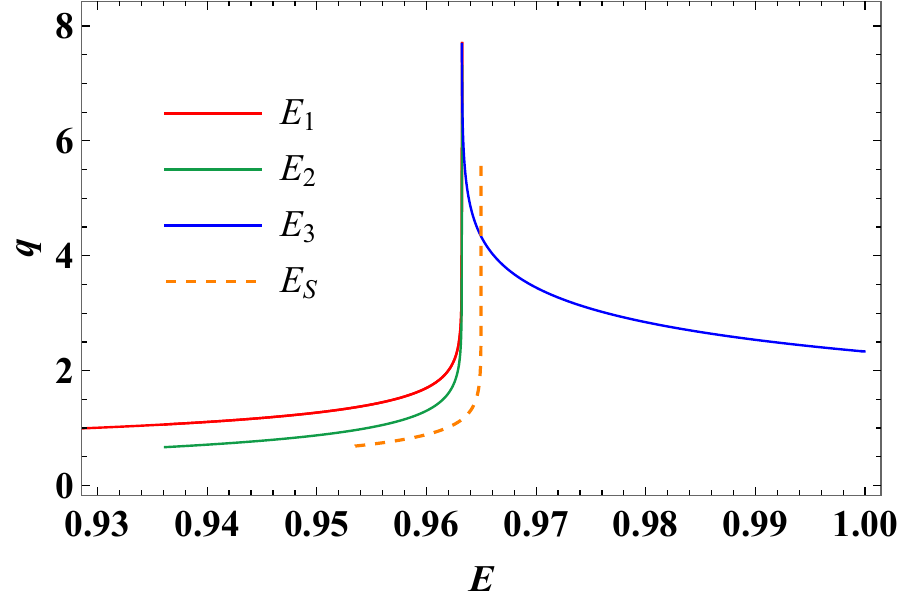}
\caption{The $E$-$q$ relation for periodic orbits at fixed angular momentum. Colored solid curves: three dyonic branches (red/green/blue) associated with separate potential wells with \(L = 3.056459521\). Orange dashed curve: Schwarzschild single-branch with \(L = 3.7\) shown as a reference.
}
\label{fig:Eq}
\end{figure}
Using Eqs.~(\ref{conservedL}) and (\ref{rdot}), we numerically compute $q$ as a function of the orbital energy, as shown in Fig.~\ref{fig:Eq}.
For the Schwarzschild black hole, $E\in(0.953514,\,0.964977)$ and $q$ increases monotonically with energy, diverging at the upper bound.
By contrast, the dyonic black hole exhibits qualitatively different behavior.
The two branches confined to individual wells ($E_1$ and $E_2$) display monotonic growth of $q$, similar to the Schwarzschild case, while the extended branch $E_3$ shows a monotonic decrease of $q$ as $E$ varies from $0.963216$ to unity.
Remarkably, the three branches overlap in $q$, implying that dynamically distinct orbits can share identical $q$ and winding number $n$.

\begin{figure*}[htbp]
\centering
\includegraphics[width=\textwidth]{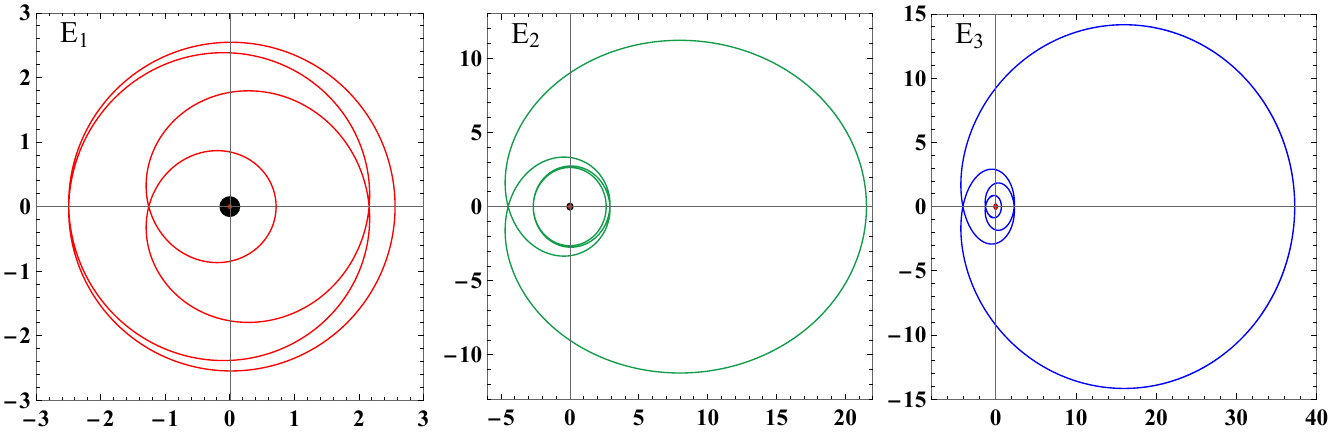}
\caption{Representative periodic trajectories with $(z,w,v)=(1,3,0)$ for the three coexisting branches $(E_1,E_2,E_3)=(0.963161895, 0.963199952, 0.976458625)$ at fixed $L=3.056459521$ and $q=3$.
All three orbits share the same winding number $n=z(1+q)$, but exhibit distinct zoom-whirl morphologies due to their localization in separate potential wells.}
\label{fig:Tra}
\end{figure*}

In Fig.~\ref{fig:Tra}, we illustrate representative periodic trajectories with $(z,w,v)=(1,3,0)$ drawn from the three coexisting branches $(E_1,E_2,E_3)$ at fixed angular momentum.
The innermost branch $E_1$ is confined to the deepest well and dominated by rapid whirl motion near the black hole horizon.
The intermediate branch $E_2$ resides in the outer well and exhibits smoother zoom-whirl behavior.
The extended branch $E_3$ spans both wells, combining features of the inner and outer families.
Despite their markedly different dynamics, all three orbits share the same winding number $n=z(1+q)=4$, confirming their topological equivalence as distinct realizations of the same periodic structure.
This topological equivalence does not imply dynamical or radiative degeneracy: owing to the multi-well structure, the three branches exhibit markedly different radial amplitudes and orbital morphologies, which, as shown below, lead to distinct gravitational-wave signatures.

{\textbf{Gravitational-wave emissions}.}---Gravitational waves emitted by an EMRI gradually extract energy and angular momentum from the smaller compact object, causing it to inspiral toward the central black hole.
For EMRIs, however, the radiative losses accumulated over a few orbital periods are typically negligible compared to the total orbital energy.
This clear separation of timescales justifies the adiabatic approximation: over a single radial-azimuthal cycle, the constants of motion $(E,L)$ remain effectively constant, and the motion of the secondary can be accurately described by geodesic dynamics~\cite{Gair:2004iv, Babak:2006uv, Hughes:2005qb}.
Within this framework, gravitational radiation governs the long-term inspiral but does not affect the instantaneous geodesic orbital morphology relevant for our analysis of multi-branch periodic motion.

Because our primary focus is the role of the background geometry, in particular, the multi-well structure of the effective potential rather than a precision treatment of radiation reaction, we adopt the standard semi-relativistic (``kludge") waveform prescription~\cite{Babak:2006uv}. In this approach, the bound geodesic trajectory in the curved dyonic spacetime is mapped onto flat-space Cartesian coordinates, and the gravitational strain is computed using the leading-order quadrupole formula, which captures the dominant contribution of the EMRI radiation. Although the quadrupole approximation is not strictly valid in the deep strong field region, it is sufficient for capturing qualitative differences in waveform morphology induced by distinct orbital dynamics.

Crucially, this approach preserves the imprint of orbital geometry zoom-whirl structure~\cite{Healy:2009zm, Wang:2025hla}, orbital radial-azimuthal frequency content, and branch dependent motion, allowing us to isolate how different effective-potential wells generate radiatively distinguishable signals even for orbits sharing identical topological indices.
Within the adiabatic approximation, the transverse-traceless metric perturbation is given by
\begin{equation}\label{GWeq}
h_{ij}^{\mathrm{TT}}(t)=\frac{2m}{D_{\mathrm{L}}}\left(a_i x_j+a_j x_i+2v_i v_j\right),
\end{equation}
where $x_i$, $v_i=\dot{x}_i$, and $a_i=\dot{v}_i$ denote the spatial position, velocity, and acceleration of the particle, respectively.
The two polarizations $(h_{+},h_{\times})$ are obtained by standard projection onto the detector frame~\cite{Babak:2006uv}.
We adopt the fiducial parameters $M=10^{7}M_{\odot}$, $m=10M_{\odot}$, luminosity distance $D_{\mathrm{L}}=200\,\mathrm{Mpc}$, latitude angle $\Phi=\pi/4$, and inclination angle $\Theta=\pi/4$.

\begin{figure*}[thbp]
\centering
\includegraphics[width=1\textwidth]{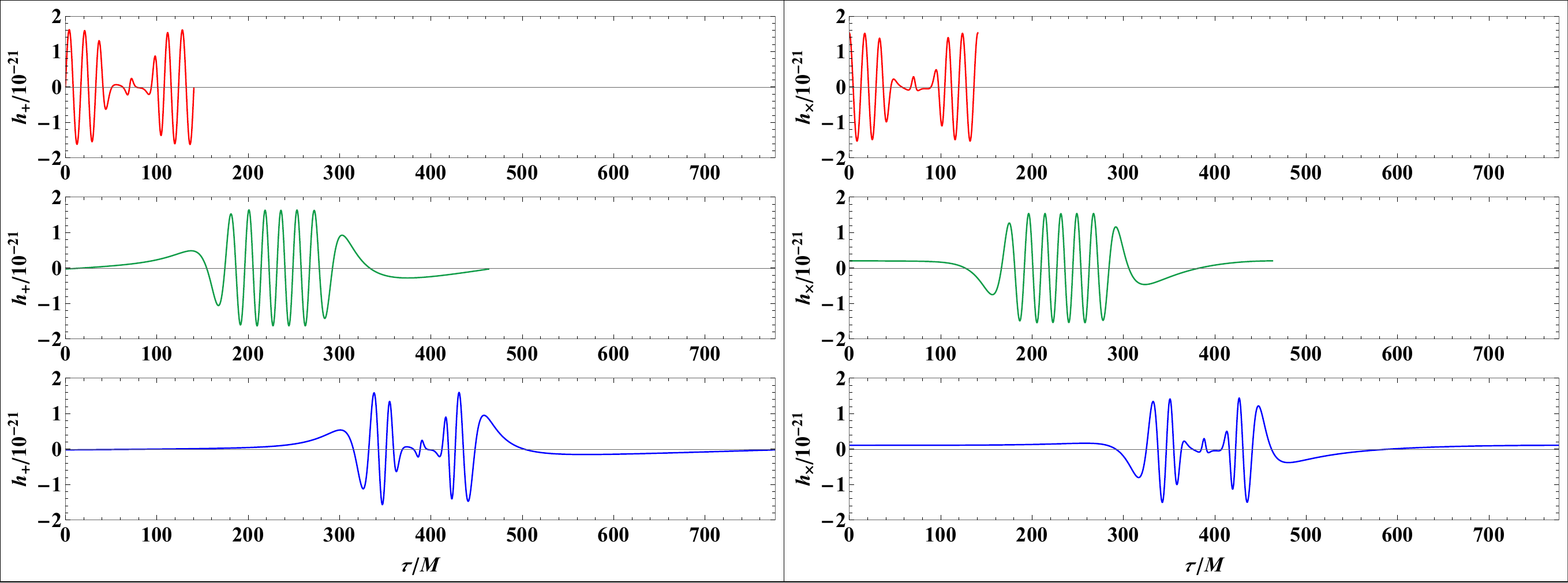}
\caption{
Plus and cross polarizations \(h_+\) and \(h_\times\) for the three branches \((E_1,E_2,E_3)\) over one orbital period.
Distinct amplitude and phase modulations encode branch dependent motion in separate potential wells.}
\label{fig:GW}
\end{figure*}

In Fig.~\ref{fig:GW}, we show the gravitational wave polarizations $h_{+}$ and $h_{\times}$ generated by the three coexisting periodic orbits belonging to the same topological class $(z,w,v)=(1,3,0)$.
For both the plus and cross polarizations, these three branches exhibit a clear amplitude hierarchy for the parameters considered here: the intermediate branch $E_2$ produces the largest signal, the inner strong field branch $E_1$ yields a smaller but still prominent amplitude, while the extended branch $E_3$ displays the weakest modulation.
Despite sharing the same rational rotation number $q$, their waveform morphologies differ markedly, reflecting the distinct zoom-whirl-vertex dynamics associated with each potential well.
The inner branch $E_1$ (red), confined to the deepest potential well, generates burst-like signals with sharp amplitude spikes near periapsis, characteristic of relativistic whirl motion in the strong field region.
The intermediate branch $E_2$ (green) exhibits smooth, quasi-sinusoidal oscillations with mild amplitude modulation during the zoom phase, closely resembling conventional EMRI waveforms~\cite{Wang:2025hla}. By contrast, the extended branch $E_3$ (blue), which spans both wells, combines features of the inner and outer families: its waveform alternates between high- and low-amplitude segments and displays phase reversals as the particle moves between these wells.

These systematic variations in amplitude and phase directly trace the multi-well structure of the effective potential, confirming that topologically equivalent orbits can produce radiatively distinct gravitational-wave signatures.
A more detailed dynamical interpretation of these waveform differences, based on the phase-space evolution of the particle’s position, velocity, and acceleration along different branches, is provided in the appendix.

\begin{figure}[htbp]
\includegraphics[width=0.47\textwidth]{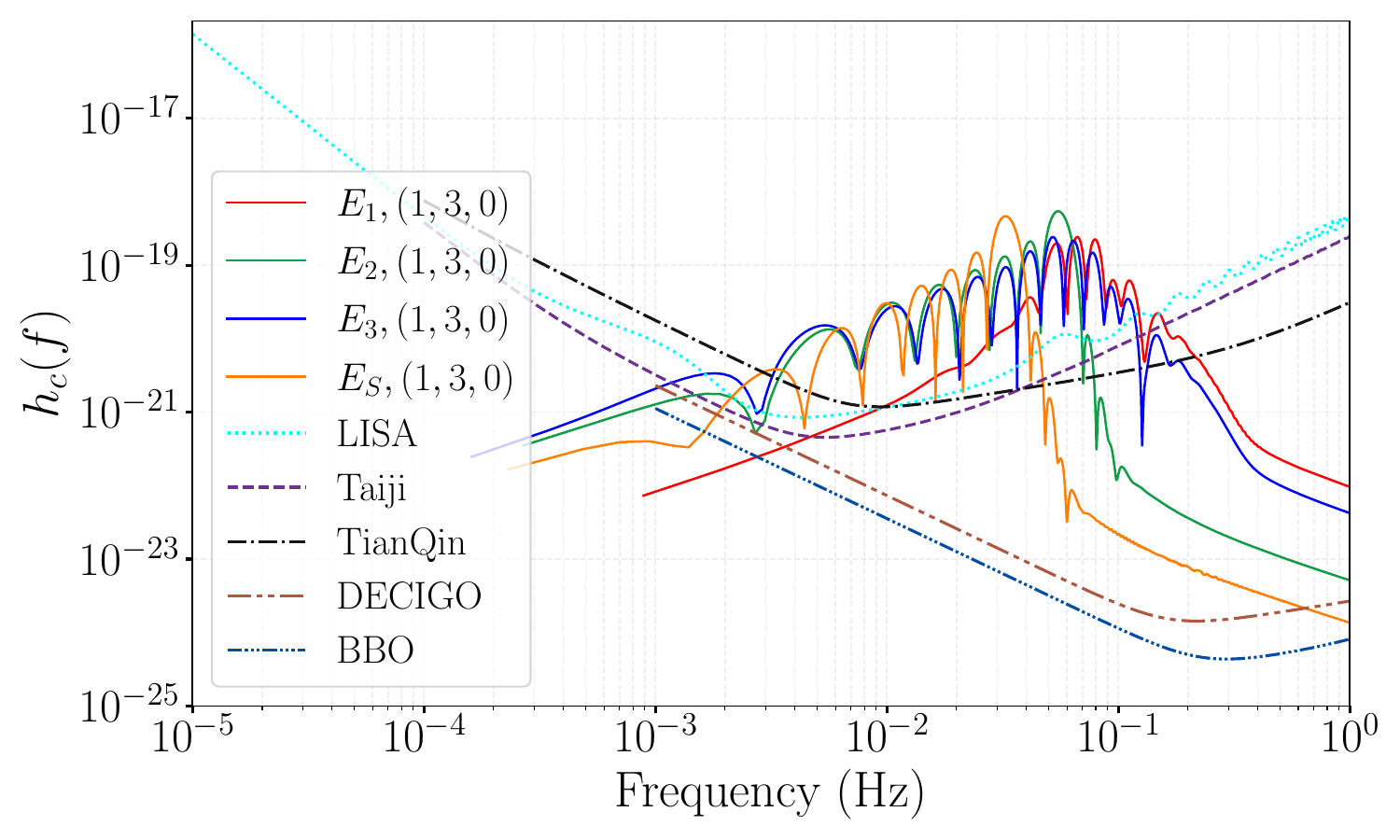}
\caption{
Characteristic strain spectra. Red/green/blue curves: dyonic branches \((E_1,E_2,E_3)\). Orange curve: Schwarzschild reference.
Sensitivity curves of space-based GW detectors are shown for comparison.
Branch dependent harmonic peaks reflect different spectral weights and frequency scalings induced by the motion of particles in distinct potential wells, despite identical rational rotation number \(q\).
}
\label{fig:strain}
\end{figure}
The corresponding characteristic strain spectra, shown in Fig.~\ref{fig:strain}, reveal clear radiative distinctions among these three branches. Although all orbits share the same rational rotation number \(q\), and thus the same ratio of the fundamental orbital frequencies \(\omega_\phi/\omega_r\), their gravitational wave spectra differ markedly. Their differences do not originate from the differences of the orbital frequency ratio, which is fixed by the common rational $q$, but from branch dependent dynamics in separate potential wells, which alter the time domain waveform morphology and redistribute the spectral weight among the gravitational wave harmonics. The inner branch $E_1$, confined to the strong field region, produces a broadband spectrum dominated by high-frequency components associated with rapid whirl motion; the outer branch $E_2$ yields a smoother, low-frequency spectrum resembling the single-well Schwarzschild case; and the extended branch $E_3$, traversing both wells, exhibits multiple frequency clusters due to the superposition of motion in the inner and outer wells. These radiative distinctions persist despite the shared topological classification $(z,w,v)=(1,3,0)$, demonstrating that topologically equivalent EMRIs can nevertheless be radiatively distinguishable as a consequence of multi-well strong field dynamics.

{\textbf{Discussion and Summary}.}---Quasi-topological nonlinear electrodynamics renders the dyonic black hole metric non-monotonic, producing multiple potential wells that host distinct families of bound periodic orbits. Remarkably, orbits sharing the same rational rotation number \(q\), and hence identical topological indices \((z,w,v)\), emit \emph{radiatively distinct} gravitational waves. By spanning different curvature regions, each branch generates characteristic amplitude modulations and harmonic structures that persist at the level of the quadrupole approximation. For fiducial EMRIs, the resulting branch dependent spectral peaks lie within the sensitivity bands of LISA~\cite{Robson:2018ifk}, Taiji~\cite{Liu:2023qap}, TianQin~\cite{TianQin:2020hid}, DECIGO, and BBO~\cite{Yagi:2011wg}, providing a concrete observational probe of the nonlinear electrodynamics in the strong field region.

These results confirm that the bound orbits of the dyonic black holes are \emph{topologically equivalent yet waveform-distinguishable}, opening a new avenue to probe strong field spacetime structure beyond the Einstein-Maxwell paradigm. While our analysis employs a semi-relativistic kludge waveform model, the reported branch dependent differences originate from the underlying geodesic structure of the multi-well spacetime and are therefore expected to persist beyond any specific waveform prescription. Future work incorporating self-force effects and Teukolsky-based waveform modeling will enable quantitative assessments of detectability and parameter estimation capabilities for these signatures.

{\textbf{Acknowledgements}.}---We would like to thank Yu-Peng Zhang and Xue-Hao Zhang for useful discussions. This work was supported by the National Natural Science Foundation of  China (Grants No. 12475055, No. 12475056, No. 12247101, and No. 12275238), the Fundamental Research Funds for the Central Universities (Grant No. lzujbky-2025-jdzx07), and the Natural Science Foundation of Gansu Province (No. 22JR5RA389, No. 25JRRA799). Tao Zhu is supported by the National Natural Science Foundation of China under Grant No. 12542053, and the Zhejiang Provincial Natural Science Foundation of China under Grants No. LR21A050001 and No. LY20A050002, and the Fundamental Research Funds for the Provincial Universities of Zhejiang in China under Grant No. RF-A2019015.

\bibliographystyle{apsrev4-2}

\appendix
\subsection*{Appendix: Gravitational-wave features of multi-branch EMRIs}\label{appedix}
To elucidate the dynamical origin \eqref{GWeq} of the waveform distinctions among different energy branches, we focus on the periodic orbit with $(z,w,v)=(1,3,0)$, which admits three coexisting branches $(E_1,E_2,E_3)$ at fixed angular momentum $L$ but distinct energies. Each branch corresponds to the motion bounded to different wells of the effective potential $V_{\mathrm{eff}}(r)$.
\begin{figure}[h]
\includegraphics[width=8.6cm]{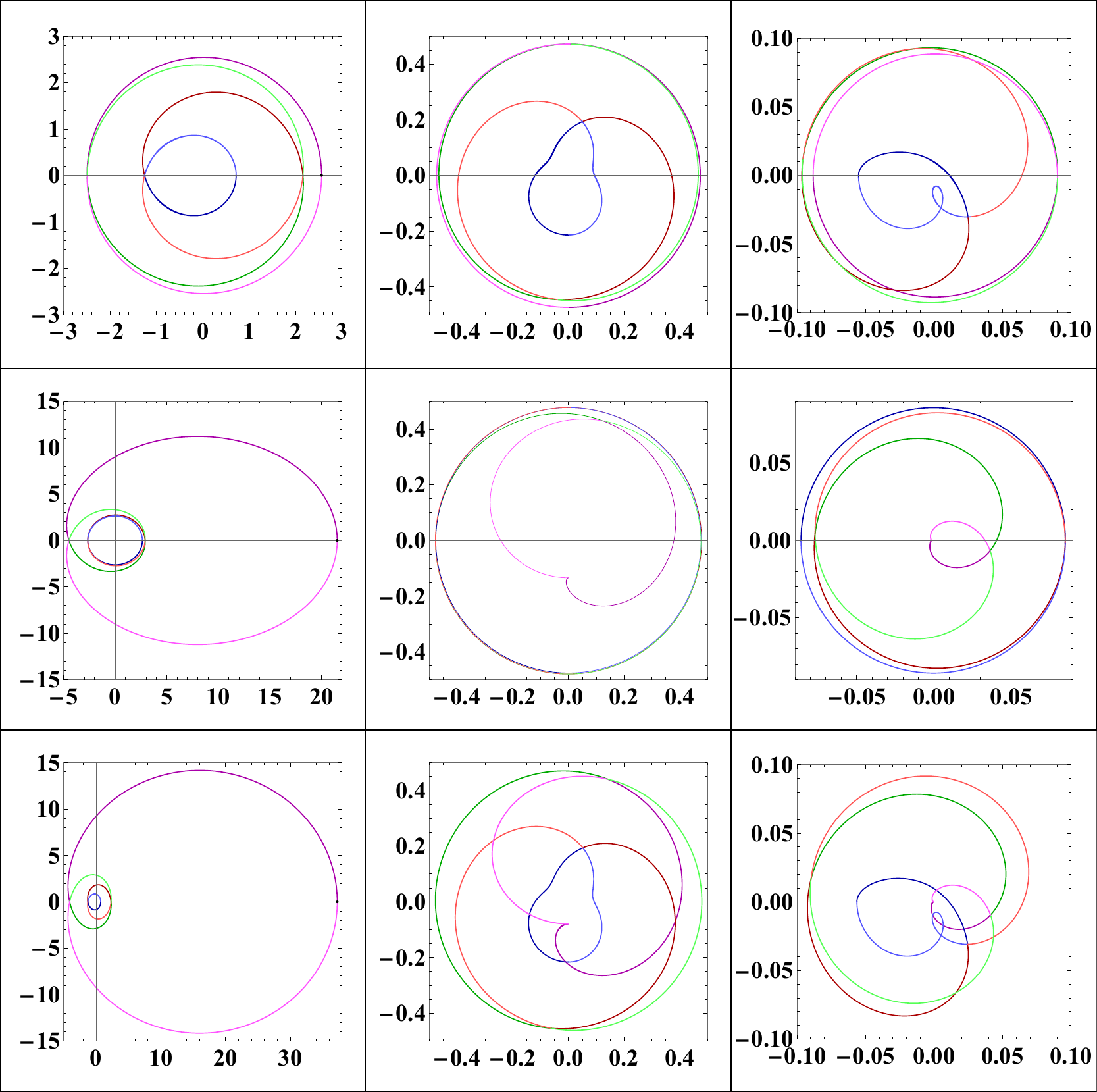}
\caption{
Phase-space evolution of the $(z,w,v)=(1,3,0)$ orbit for the three energy branches $(E_1,E_2,E_3)$.
Columns display the position $x$, velocity $v$, and acceleration $a$; rows correspond to these three branches.
Each trajectory covers one complete radial cycle.}
\label{fig:xva}
\end{figure}

As shown in Fig.~\ref{fig:xva}, the $E_1$ branch remains bounded to the innermost potential well and exhibits smooth, nearly circular motion with relatively small velocity and acceleration amplitudes.
The $E_2$ branch, localized in the outer well, displays broader excursions and a mild nonlinear coupling between radial and azimuthal motion. By contrast, the highest energy branch $E_3$ traverses both wells, giving rise to multi-lobed and self-intersecting velocity patterns, accompanied by sharp acceleration bursts near the turning points.
In short, these features reveal a clear dynamical progression from regular single well oscillations to strongly nonlinear, cross well motion as the orbital energy increases.

\begin{figure}[htbp]
\includegraphics[width=8.6cm]{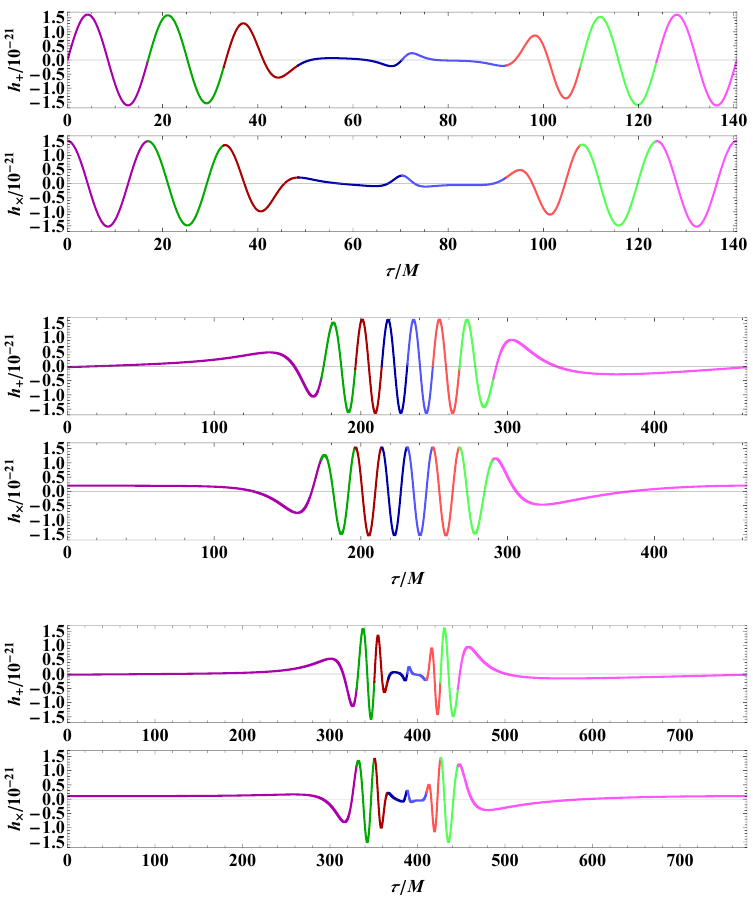}
\caption{
Gravitational waveforms $h_{+}$ and $h_{\times}$ for the three energy branches $(E_1,E_2,E_3)$ of the $(z,w,v)=(1,3,0)$ orbit.
Despite identical topological classification and angular momentum, the waveform morphology varies significantly with orbital energy, reflecting the multi-well structure of the effective potential.}
\label{fig:pieceGW}
\end{figure}

These dynamical differences manifest directly in the gravitational waveforms shown in Fig.~\ref{fig:pieceGW}, computed within the quadrupole approximation for an EMRI system. The $E_1$ branch produces an approximately sinusoidal, narrow band signal.
The $E_2$ branch exhibits smoother amplitude modulation with moderate variability. In contrast, the $E_3$ branch generates broadband, burst-like emission, characterized by pronounced amplitude spikes and phase distortions as the particle transitions between these potential wells.

Overall, the sequence from $E_1$ to $E_3$ represents a transition from stable, quasi-harmonic radiation to highly modulated and strongly nonlinear emission.
These waveform features directly reflect the underlying multi-well structure of the dyonic spacetime and provide complementary evidence for the radiative distinguishability of topologically equivalent EMRIs.
Such signatures may offer an additional observational handle for identifying multi-branch dynamics in future space-based gravitational wave missions.

\end{document}